\documentclass[11pt,a4]{article}

\usepackage{amssymb,graphicx,amsmath,sectsty,MnSymbol}
\sectionfont{\large\bf}
\subsectionfont{\large\it}

\graphicspath{{figs/}}
\long\def\symbolfootnote[#1]#2{\begingroup%
\def\thefootnote{\fnsymbol{footnote}}\footnote[#1]{#2}\endgroup}

\begin{document}

\begin{center}

\LARGE {\bf  Simulations of the Nuclear Recoil Head-Tail Signature in Gases Relevant to Directional Dark Matter Searches}

\vspace{1cm}

\Large  P.~Majewski$^{a,}$\symbolfootnote[1]{Corresponding author:  \\ {\it Email address:} {\tt pawel.majewski@stfc.ac.uk} (P.Majewski)},
D.~Muna$^a$, D.P.~Snowden-Ifft$^b$, N.J.C.~Spooner$^a$

\normalsize
\vspace{.2cm}

{\noindent \it $^a$ Department of Physics and Astronomy, University of Sheffield, S3 7RH, UK }\\
{\noindent \it $^b$  Department of Physics, Occidental College, Los Angeles, CA 90041, USA}

\end{center}

\begin{center}
\begin{tabular*}{1\textwidth}{c}
\hline
\end{tabular*}
\end{center}

\section*{Abstract} 
{\small
We present the first detailed simulations of the head-tail effect relevant to directional 
Dark Matter searches. Investigations of the location of the majority of the ionization 
charge as being either at the beginning half (tail) or at the end half (head) of 
the nuclear recoil track were performed for carbon and sulphur recoils in 40 Torr 
negative ion carbon disulfide and for fluorine recoils in 100 Torr carbon tetrafluoride. 
The SRIM simulation program was used, together with a purpose-written Monte 
Carlo generator, to model production of ionizing pairs, diffusion and basic readout 
geometries relevant to potential real detector scenarios, such as under development 
for the DRIFT experiment. The results clearly indicate the existence of a 
head-tail track asymmetry but with a magnitude critically influenced by two competing 
factors: the nature of the stopping power and details of the range straggling. The 
former tends to result the tail being greater than the head and the latter the reverse.  
}
\vspace{0.8cm}

\noindent {\it Key words:} Dark matter, Directional detector, Gas detector, WIMP, neutralino, 
TPC, Simulation \\
\noindent {\it PACS:} 95.35.+d, 29.40.Cs42, 87.55.Gh 

\begin{center}
\begin{tabular*}{1\textwidth}{c}
\hline
\end{tabular*}
\end{center}

\vspace{0.4cm}

\section{Introduction}

Measurement of the direction of a low energy ($\approx$1~keV/amu) nuclear 
recoil track and the ionization charge distribution along it, resulting from the 
elastic scattering of a target nucleus by an incoming WIMP 
(Weakly Interacting Massive Particle) provides, an unambiguous identification of WIMPs 
as responsible for the galactic dark matter \cite{ref1}. Amongst current radiation 
detection technologies only Time Projection Chambers filled with low pressure 
gas, appear capable of such a measurement. This prospect has been 
demonstrated by the directional Dark Matter search R\&D experiments DRIFT \cite{ref2}, 
NEWAGE \cite{ref3} and MIMAC \cite{ref4} using Time Projection Chambers (TPCs) filled 
with low pressure CS$_2$ and CF$_4$. In these detectors attempts are currently 
made to reconstruct the orientation of the low energy nuclear recoil tracks, 
typically of a few millimetres in length. However, due to the character of the 
electronic and nuclear stopping powers of low energy nuclear recoils in gas, an 
asymmetric ionization charge distribution along their tracks may also be expected. 
Thus additional information on the absolute direction of recoils might 
be also available. Such potential information on the track sense is termed the 
head-tail effect, for instance describing the location of the majority of the 
ionization charge as being either at the beginning half (tail) or at the end half 
(head) of the track. It is known that if this information can clearly be extracted 
in a detector this would, by breaking the forward-back degeneracy in 
the direction of recoils from WIMPs, have a dramatic impact on the potential 
directional sensitivity to dark matter, an effect likely to be of order $\times$10 or 
more \cite{ref5}.

Realisation of this gain would greatly increase the feasibility of building a large 
dedicated low pressure TPC detector capable of a definitive identification of 
dark matter. It is thus vital to understand whether, and to what magnitude, 
the head-tail effect is present, in both principle and practice. Recently some 
experimental evidence for head-tail asymmetry has been observed for F recoils 
in 100-380~Torr CF$_4$ \cite{ref6}. Observation of the effect there was possible due to 
the use of relatively high energy recoils ($>$200~keV) and a short drift gap. 
However, measurement of the effect in a more realistic scenario for dark matter 
searches, in a large volume (order m$^3$), with negative ion gas and at lower 
recoil energy, is more difficult. This is due to the limited spatial resolution 
imposed by the current readout technologies used for larger scale detectors, 
for instance Multiwire Proportional Counters (MWPCs) as in DRIFT, and 
due to the inevitable effects of ionization charge diffusion in large detectors. 
Only preliminary experimental measurements have been possible so far with 
DRIFT itself \cite{ref7}. Thus, in view of the importance of the issue for directional 
dark matter detection, the need to understand better the physical processes 
so as to help optimise implementation of possible head-tail discrimination in 
a large detector, it is useful to undertake detailed simulations of the head-tail 
response and start to make theoretical predictions. This is the objective of the 
work presented here. In this light it should be noted that the purpose here is 
not to detail a complete head-tail simulation for a specific detector such as 
DRIFT II (the subject of a separate paper) but to explore the issue in a more 
generic fashion. 

So far some basic theoretical predictions of the head-tail effect in binary gases 
have been performed by A. Hitachi, i.e. \cite{ref8}. These results were based on 
the Linear Energy Transfer calculations and ion projected range estimations 
from SRIM \cite{ref9}. The obtained Bragg-like curves indicated that more ionisation 
charge is produced at the tail than at the head of the track, regardless of the 
ion type or energy. However, that work did not attempt to account for straggling, 
or address the issue of diffusion and the influence of track reconstruction 
geometry. In this work we present first detailed results of simulations of the 
energy loss and ionization charge distribution along tracks of carbon and sulfur 
ions in 40~Torr CS$_2$ ($\rho$=$1.67\cdot10^{-4}$~gcm$^{-3}$)
and for fluorine ions in 100 Torr 
CF4 ($\rho$=$4.73\cdot10^{-4}$gcm$^{-3}$). 
The head-tail effect is studied as a function of ion 
energy and diffusion but also in relation to the effect of straggling and the 
dependence on readout geometry. 

\section{Simulation procedures and SRIM results}

The basis of this work is firstly the SRIM (Stopping and Range of Ions in 
Matter) software package and its incorporated TRIM track generator Monte 
Carlo program. SRIM is a code commonly, and primarily, used for predictions 
of parameters relevant to ion implantation, sputtering and transportation in 
solid materials. However, it is also successfully used to simulate interactions 
in gases, though for that application there has been less experimental 
verification so that some caution is needed, particularly at the low pressures relevant 
to this work. SRIM is used here to produce tables as a function of energy of 
both the electronic (S$_e$) and nuclear (S$_n$) energy loss, stopping power, range, 
longitudinal and lateral straggling of nuclear recoils (ions) in the gaseous 
target chosen. The TRIM component is then used to generate many individual 
simulated tracks. This allows initial investigation of the energy loss 
distribution along individual raw recoil ion tracks and their secondaries, on an 
event-by-event basis. To complete the process, to give relevance to real detector
 scenarios, it is then necessary to convert, using the known average energy 
to create electron-ion pairs (W-value), the energy loss along each recoil track 
into ionization charges, or number of Negative Ion Pairs (NIPs), distributed 
along the track. Finally, it is necessary to account for the effects of diffusion 
of the tracks through a realistic drift volume (taken here to be up to 50~cm) 
and of the effects of projection onto readout axes. An in-house Monte Carlo 
was written for these latter stages, with appropriate values for the W to NIPs 
conversion factor, changing as a function of ion energy. 

\begin{figure}
\begin{center}
\includegraphics{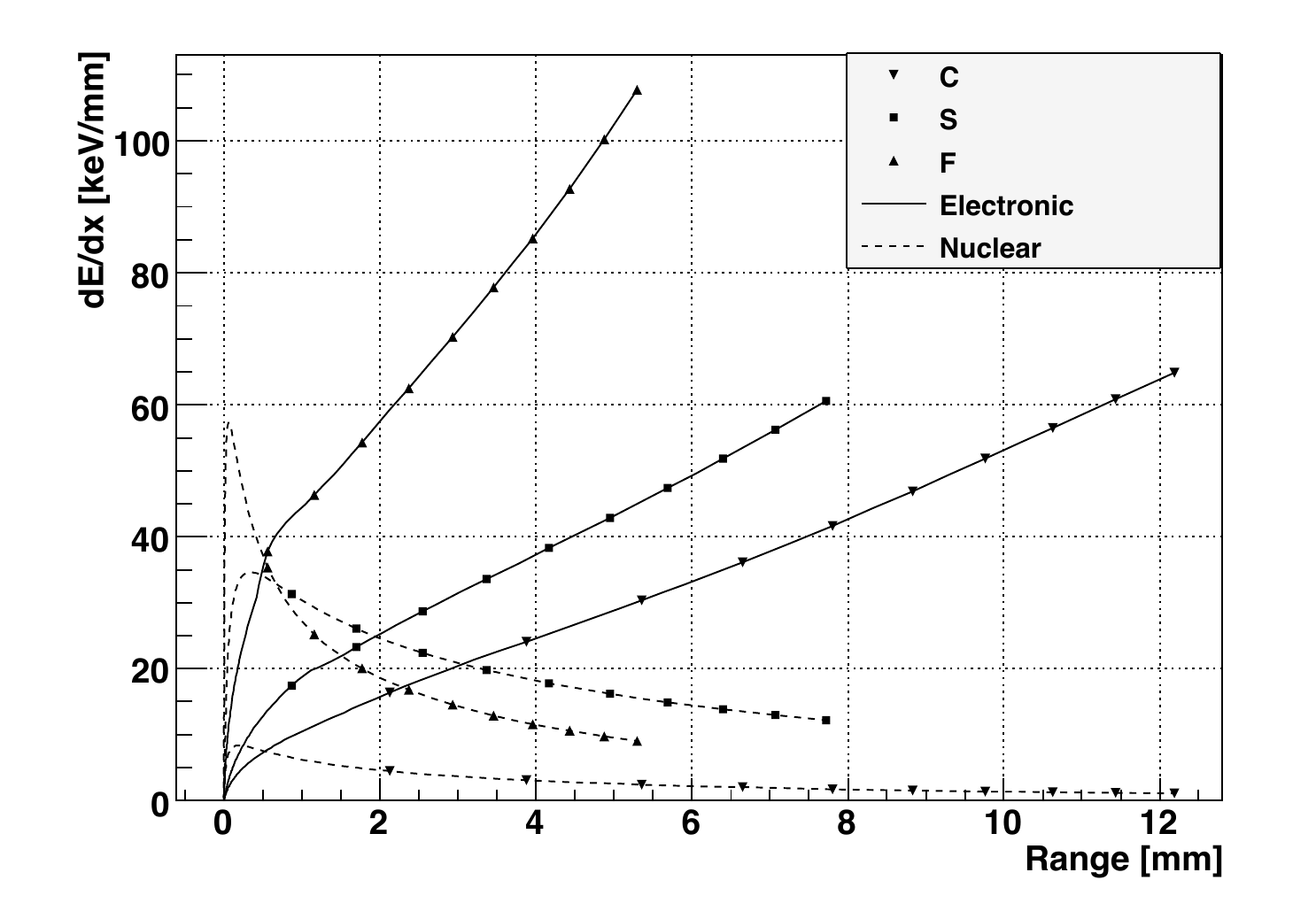}
\caption{Calculated with SRIM ionization energy loss dE/dx as a function of ion 
range in the medium. Energy losses in electronic (solid line) and nuclear (dashed 
line) channels are shown for Sulfur ($\blacksquare$) and Carbon ($\filledmedtriangledown$) ions in 40~Torr CS$_2$ and 
Fluorine ($\filledmedtriangleup$) in 100~Torr CF$_4$. Ion energy is marked up to 500~keV with an interval 
of 50~keV. }
\label{fig1}
\end{center}
\end{figure}

\begin{figure}
\begin{center}
\includegraphics{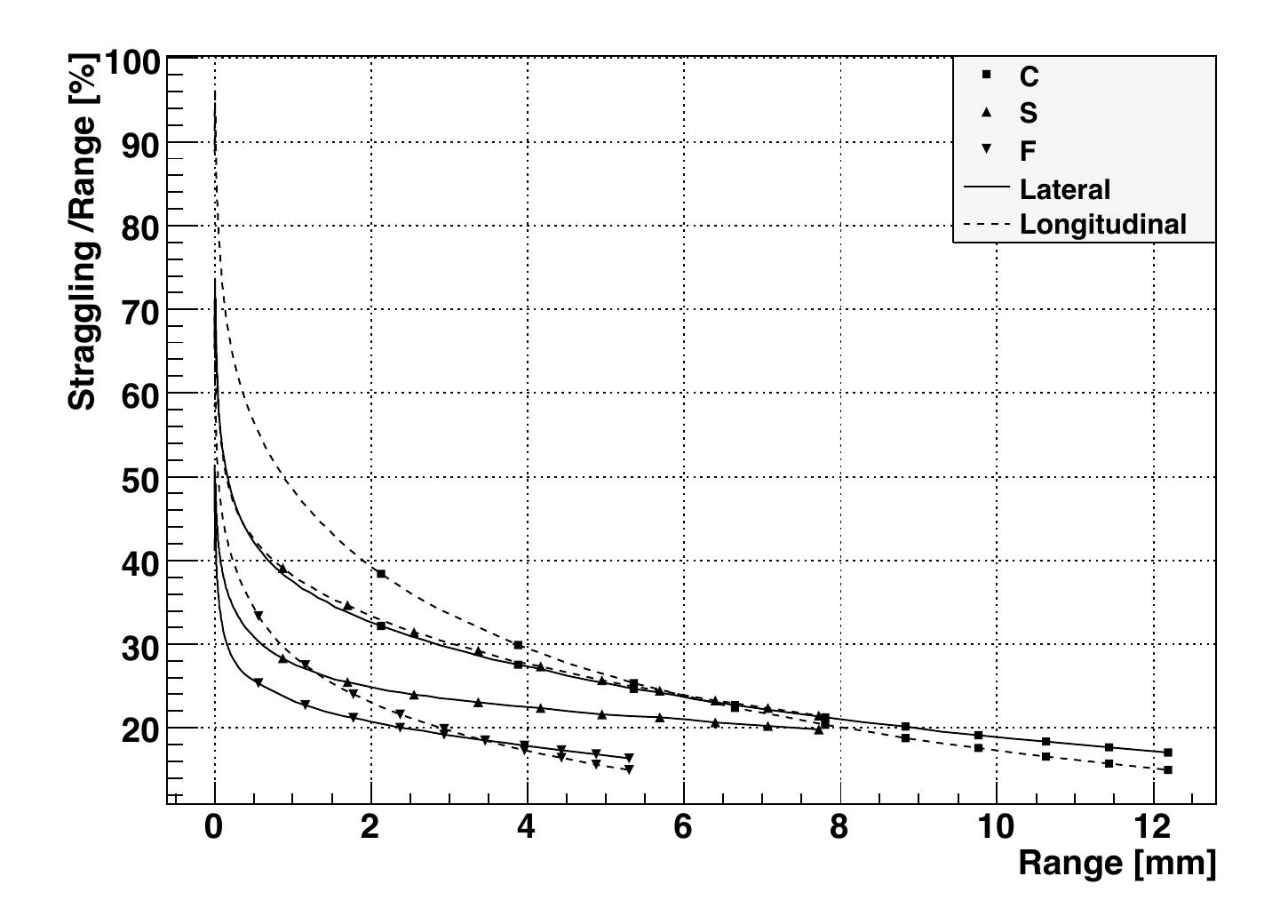}
\caption{Lateral (solid line) and longitudinal (dashed line) straggling/range as a 
function of range for Fluorine ions in CF$_4$ ($\filledmedtriangledown$), Sulfur ($\filledmedtriangleup$) and Carbon ($\blacksquare$) ions in 
 CS$_2$ which energy is marked up to 500~keV with an interval of 50~keV. }
\label{fig2}
\end{center}
\end{figure}

In this work we calculate parameters for the selected recoil ions using energies 
up to 500~keV. Fig. \ref{fig1} and Fig. \ref{fig2} 
show SRIM results for the energy loss, lateral 
and longitudinal straggling, plotted against recoil ion range for ion energies 
selected at 50~keV intervals. As can be seen in these figures, at higher energies 
the electronic energy loss is dominant and decreases with energy, whereas at 
lower values the nuclear energy loss becomes greater than the electronic and 
increases with decreasing energy, at least initially. For example, for sulphur 
ions the nuclear energy loss starts to dominate at~100 keV. This means that 
the energy loss along the ion track is not continuously decreasing, as might 
naively be expected. 

\begin{figure}
\begin{center}
\includegraphics{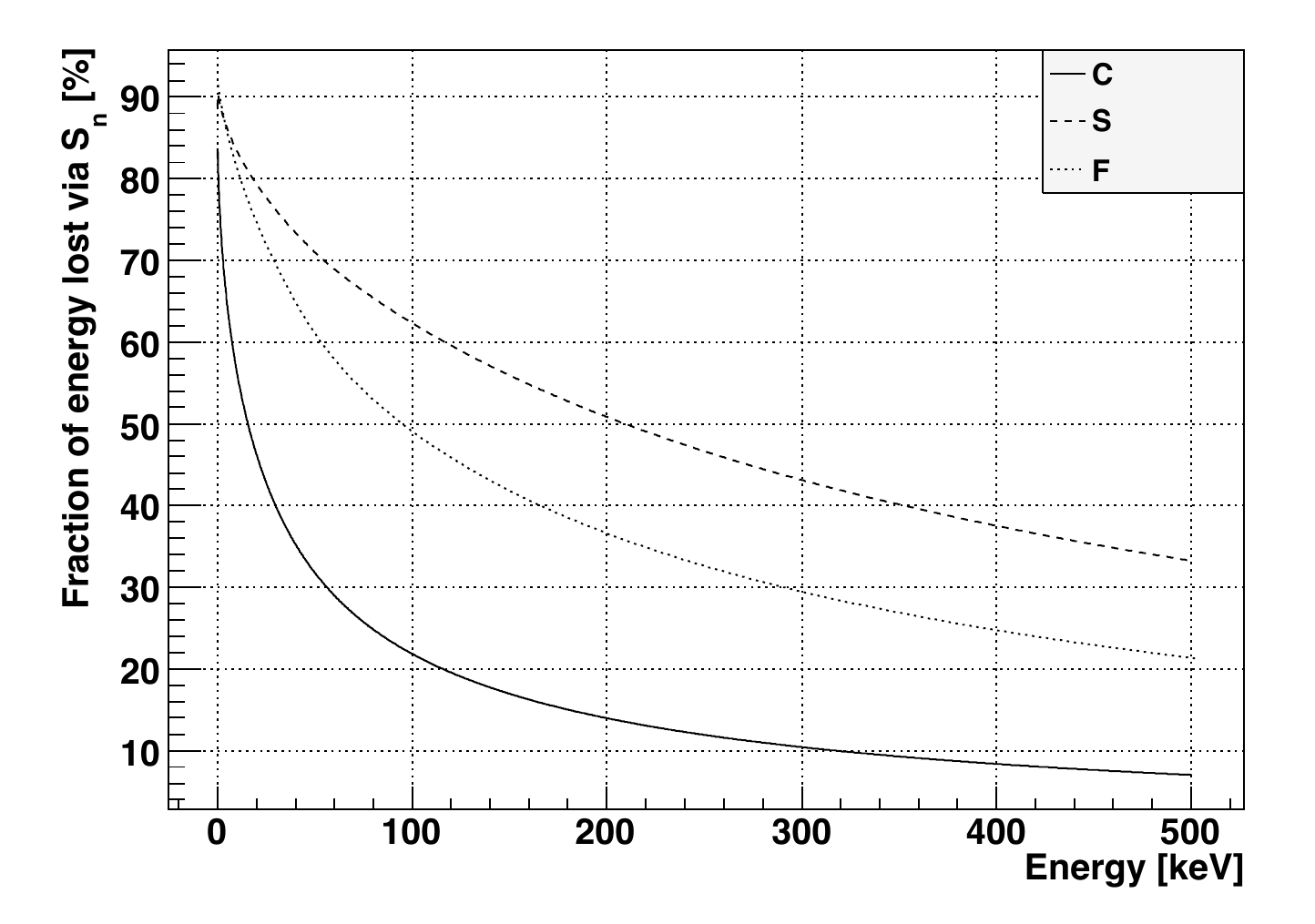}
\caption{Energy loss due to nuclear interactions with respect to the total energy loss 
for Carbon, Sulfur in CS$_2$ and Fluorine ions in CF$_4$.}
\label{fig3}
\end{center}
\end{figure}

The amount of energy loss due to nuclear interactions as a function of ion 
energy is presented in Fig \ref{fig3}. The difference in nuclear energy loss between 
the different ions is clearly seen here. For carbon ions 50\% or more of the 
total energy is lost via nuclear interactions for energies E$<$20~keV, whereas in 
fluorine and sulphur the equivalent factor occurs at much higher energy, 90 
and 200~keV, respectively. This energy loss produces very low energy nuclear 
recoils which can produce significant ionization. 

Due to continuous collisions with gas atoms the direction of moving ions also 
deviates from their original path. This causes fluctuations in the recoil ionÕs 
range described by the lateral and longitudinal straggling as a function of ion 
energy. This parameter, divided by the range, is shown in Fig \ref{fig2}. Here one can 
see that as the ions slow down they experience increasing straggling relative 
to the drift length they have ahead, rising sharply at the very end of the track. 

Tables of parameters such as those above, calculated using SRIM, are used 
in the Monte Carlo generator component of SRIM called TRIM to generate 
tracks. In addition to on-line histogramming of various quantities, TRIM 
results are also recorded event by event in two different formats. The first is 
included in the file called {\it COLLISON.TXT}, generated either in quick 
simulation mode or as a full nuclear recoil cascade simulation. For the former mode 
the current energy, position in 3D and type of secondary recoils with their 
initial energy is recorded for every ion collision. For the latter, information 
about the energies and positions of all created nuclear recoils is added. TRIM 
also creates an output file called {\it EXYZ} which delivers information on the 
primary ion electronic and nuclear energy loss along the track with high spatial 
resolution. This format was chosen here for analysis of the head-tail effect. 

On this basis, for each recoil ion with a given energy, 10$^4$ events were generated, 
recorded and further analysed. The energies used were for fluorine: from 100 
to 400 with a step of 50~keV; sulphur: from 10 to 100 with a step of 10~keV 
then up to 250 with a step of 25~keV and up to 400 with a step of 50~keV and 
carbon from 10 to 100 with a step of 10~keV then up to 200~keV with a step 
of 25~keV and up to 300 with a step of 50~keV. Each track event in the TRIM 
output {\it EXYZ} file contained on average 100 collision points with information 
on the current ion energy, the 3D coordinate of the collision point and the 
electronic energy loss. 

In order to generate the number of electron-ion pairs, produced along a recoil 
track it is essential to know the W-value and its dependence on ion energy. 
In this way, for each ion of a given energy, an appropriate W-value can be 
selected at every collision point according to its current energy.

\begin{figure}
\begin{center}
\includegraphics{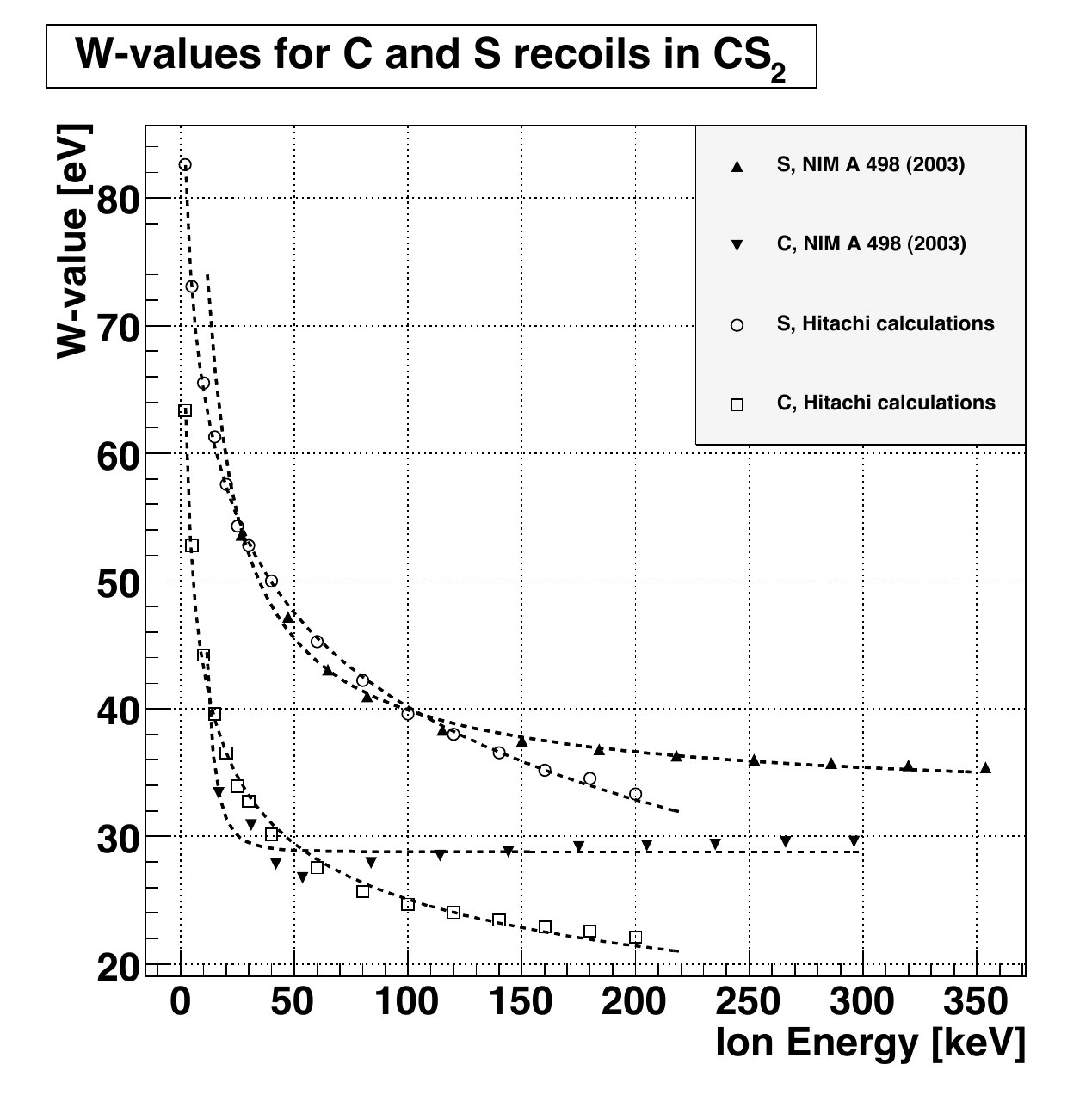}
\caption{Carbon and Sulfur ions W-values in CS$_2$ gas as a function of energy. Points 
marked with ($\filledmedtriangleup$ and $\filledmedtriangledown$) show results from \cite{ref10} while ($\medsquare$ and $\medcircle$) values calculated 
by A. Hitachi. Dashed lines are from a Þt used for {\it EXYZ} TRIM output file analysis. }
\label{fig4}
\end{center}
\end{figure}

For carbon and sulphur ions W-values were selected using the data shown in 
Fig. \ref{fig4}. Here two sets of values are presented, the first set are measured values 
taken from \cite{ref10} and the second are predicted values derived by A.Hitachi \cite{ref11}. 
As seen in Fig. \ref{fig4} there is only partial agreement between published data and 
simulation. Furthermore, in both cases the value of W at energy below 20~keV 
is not well determined. For the purposes of this work, as shown in Fig. \ref{fig4}, it 
was decided to use primarily the experimental data and to use a fit function 
in the form of: $f(x) = f (x) = a + \frac{b}{x^c}$ where a, b and c are free parameters, to the data 
points, including an extrapolation below 20~keV. Values of the fit parameters 
used in this work are shown in Table \ref{tab2}. 

\begin{figure}
\begin{center}
\includegraphics{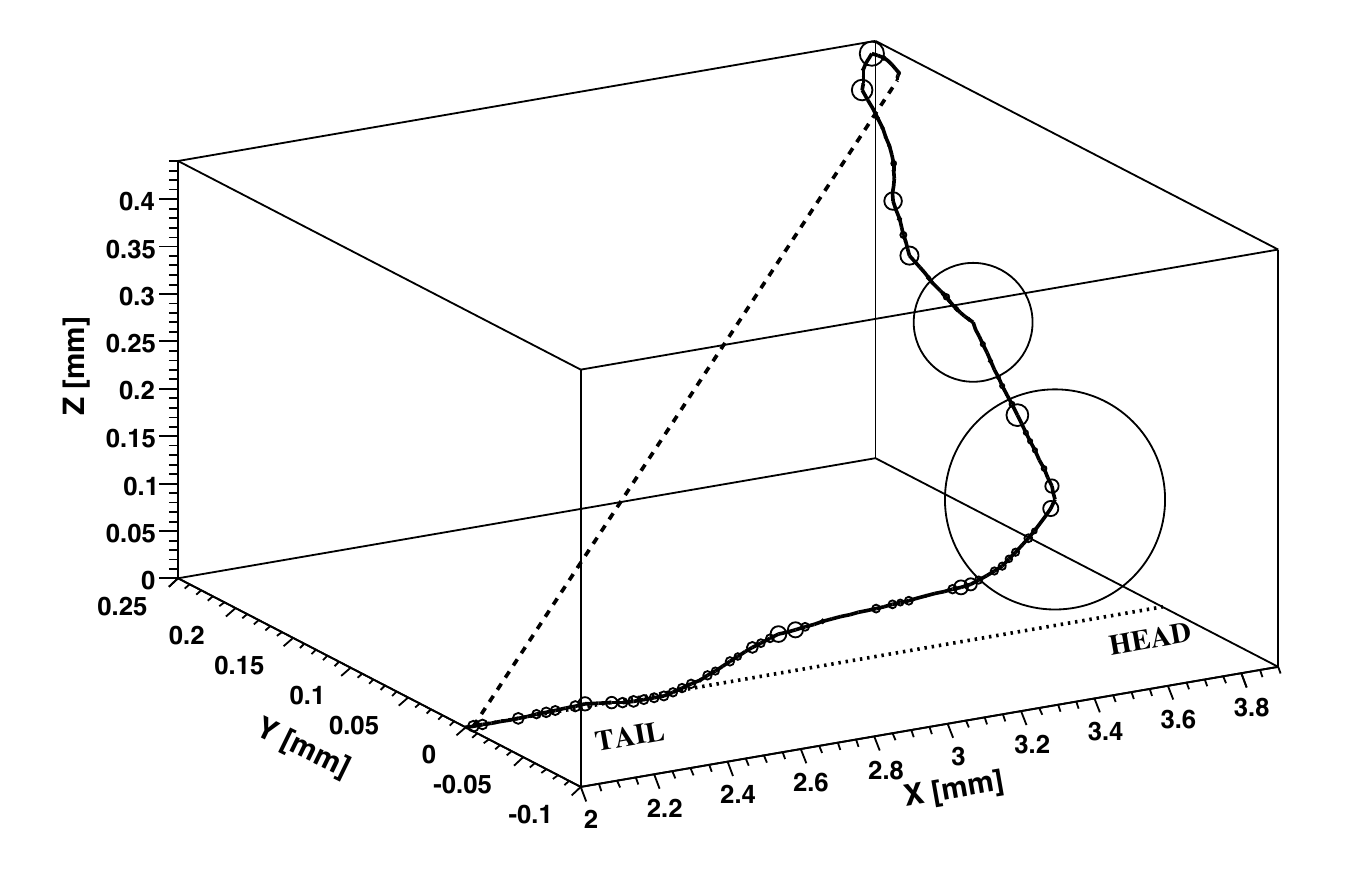}
\caption{A single track of 100~keV Sulfur ion obtained from {\it EXYZ} TRIM output 
Þle. In the work the ionization energy spatial distribution and head-tail effect were 
studied along: ion track line (solid) and along the axis of the initial motion (dotted). 
The projected range is denoted by the dashed line. Size of the circles along the track 
is proportional to the magnitude of the energy loss in the collision. Tail and Head 
show the position of the beginning and the end of the track, respectively. }
\label{fig5}
\end{center}
\end{figure}

Using the W-value fit of Fig. \ref{fig4}, the ionisation energy loss and ionisation charge 
spatial distributions, and hence head-tail effects, were studied for two different 
geometric configurations relative to the ion interactions points: (i) along the 
track, and (ii) projected onto an axis aligned with the initial direction of the 
ion. These geometries allow exploration of the influence of readout on head-tail 
sensitivity. In particular, first configuration represents the ideal case of perfect 
3D track reconstruction. The second configuration is relevant to a simpler 1D 
type readout. Fig. \ref{fig5} shows an illustrative plot with projected lines for readout 
geometries for an example 100~keV sulphur ion track. On this plot the small 
circles along the track represent interaction points of energy loss with the size 
being proportional to the magnitude of the energy loss. It is important to 
consider that different range straggling will occur for ions of the same energy. 
Hence, for comparison purposes in this plot and elsewhere, the position of each 
interaction projected point has been normalised to the projected range of the 
ion. 

\subsection{Ion Ranges}

\begin{figure}
\begin{center}
\includegraphics{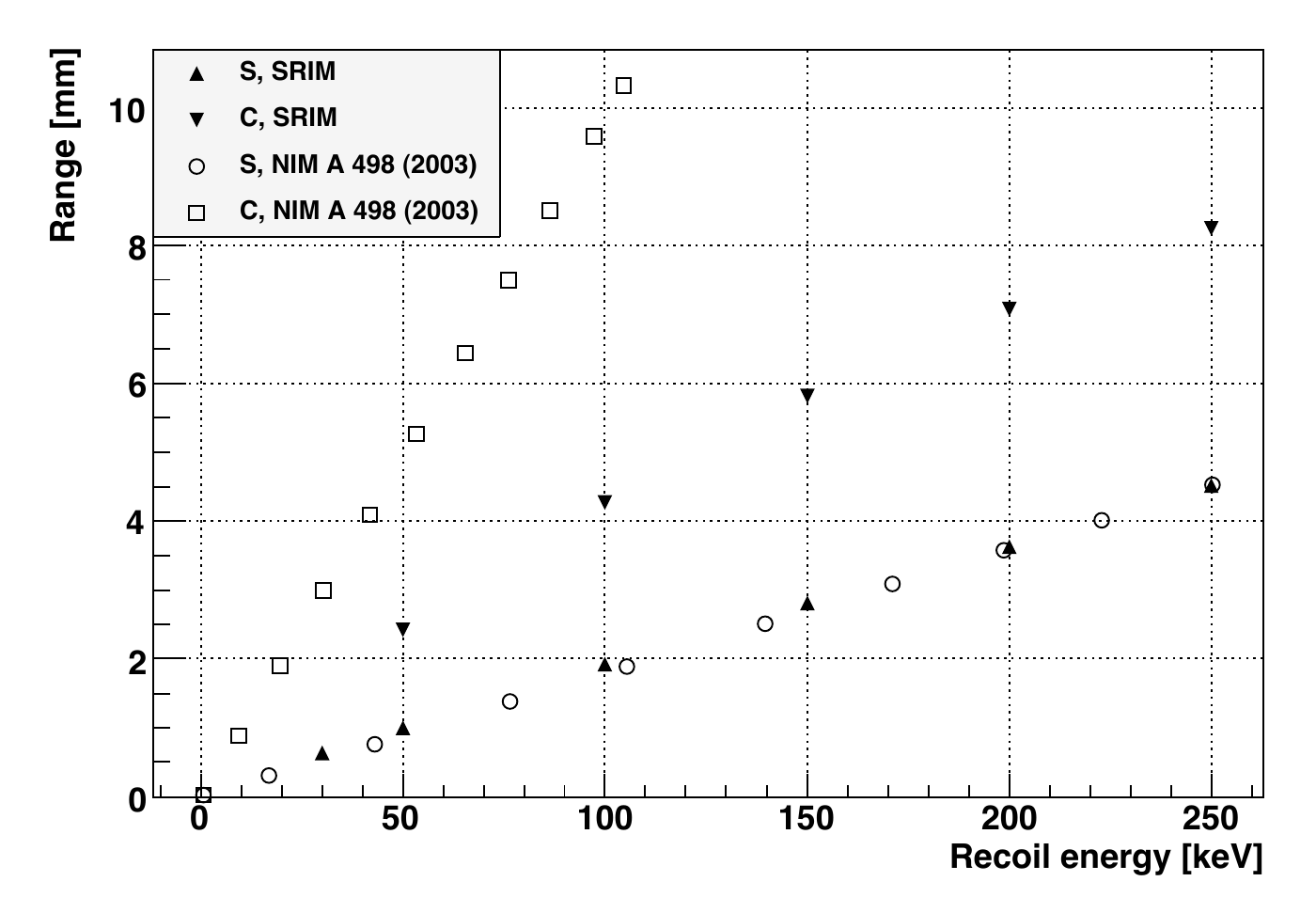}
\caption{Sulfur and Carbon ion range in CS$_2$ as a function of the energy, calculated 
with SRIM ($\filledmedtriangleup$ and $\filledmedtriangledown$) and taken from \cite{ref10} ($\medcircle$ and $\medsquare$). }
\label{fig6}
\end{center}
\end{figure}

\begin{figure}
\begin{center}
\includegraphics{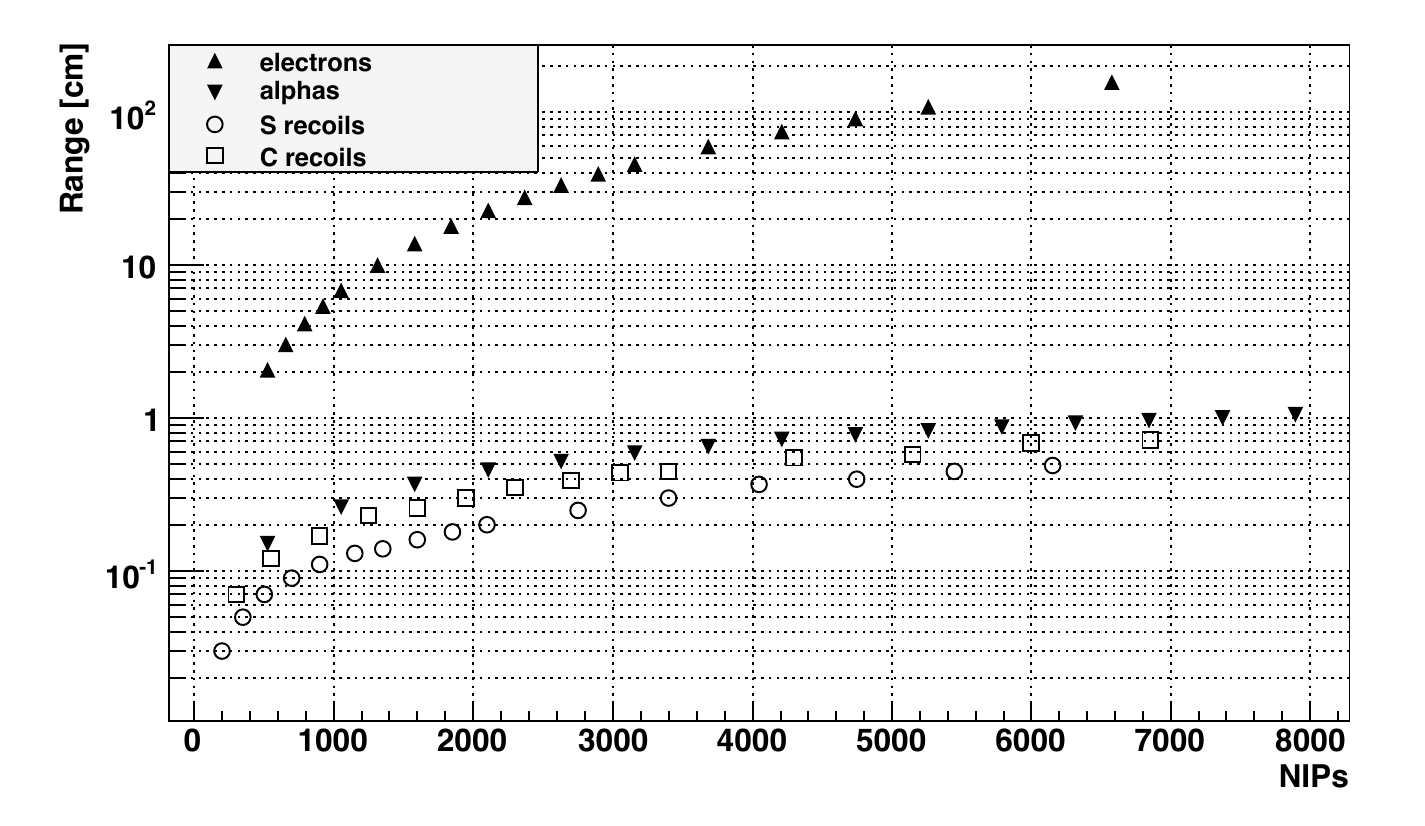}
\caption{Sulfur ($\medcircle$), Carbon ($\medsquare$), alpha ($\filledmedtriangledown$) and 
electron ($\filledmedtriangleup$) range in CS$_2$ as a 
function of ionization charge (NIPs) created. NIPs for Sulfur and Carbon were 
calculated using the W-values shown inFig. \ref{fig4} while for alpha particles and electrons 
a constant value of W=19~eV was used. To calculate the range of electrons ESTAR 
program was used for energies from 10 to 125~keV with steps 2.5, 5 and 10 up to 20, 
60 and 100~keV, respectively. Results for alpha particle ranges are for energies from 
10 to 150 with a step of 10~keV. }
\label{fig7}
\end{center}
\end{figure}

Based on the processes described above a study was performed first to 
compare ion ranges, before diffusion, as a function of energy and NIPs. Fig. \ref{fig6}
shows results for sulphur and carbon in CS$_2$ vs. energy in comparison with 
data from \cite{ref10}. Here, the projected range was used, calculated using the 
coordinates of the first and last point of the ion track. It can be seen that there 
is a very good agreement for sulphur. For carbon ions however, agreement is 
less compelling, the results from this work yielding less than half the values 
in \cite{ref10}. This discrepancy, which is the subject of further experimental study, 
although significant does not effect the general conclusions of this work, since 
sulphur and fluorine have the dominant response for dark matter scattering. 
For the conversion of the energy loss into ionization charge, the number of 
electrons attached, with no loss, to CS$_2$ molecules creating CS$_2-$ 
was summed 
for each carbon and sulphur ion energy. Fig. \ref{fig7} shows, as a function of NIPs, the 
resulting range of nuclear recoils together with the range of alpha particles and 
electrons calculated with TRIM and ESTAR \cite{ref12}, respectively. The NIP values 
were calculated here for the latter species using a W-value of 19~eV \cite{ref13}. It can 
be seen from Fig. \ref{fig7}, comparing electron recoils with nuclear recoils yielding 
the same number of NIPs, that the difference in range is typically one order 
of magnitude greater for electrons than for nuclear recoils. This illustrates the 
expected great electron (gamma) rejection power of a TPC operating with 40~Torr 
CS$_2$, as has been experimentally confirmed \cite{ref14}. However, it also shows 
that the range of nuclear recoils and alpha particles of similar low energy are 
close to each other, the difference being not greater than 1-2~millimeters for 
tracks a few millimetres long.

To check further the consistency of the SRIM procedure used here, and de- 
veloped for the head-tail analysis below, some comparisons was made with 
existing measurements. In particular, the ranges of 103~keV $^{206}$Pb recoils in 
several pure and binary gases at STP were calculated and compared with data 
from \cite{ref15}. The results are shown in Table \ref{tab1}. As can be seen in most cases there 
is less than a 10\% difference. Similar differences between measurements and 
simulation were observed for long range 5-6~MeV alpha particles from Rn and 
Po decays in 40~Torr CS$_2$ \cite{ref16}. 

\subsection{Head-Tail}

\begin{figure}
\begin{center}
\includegraphics{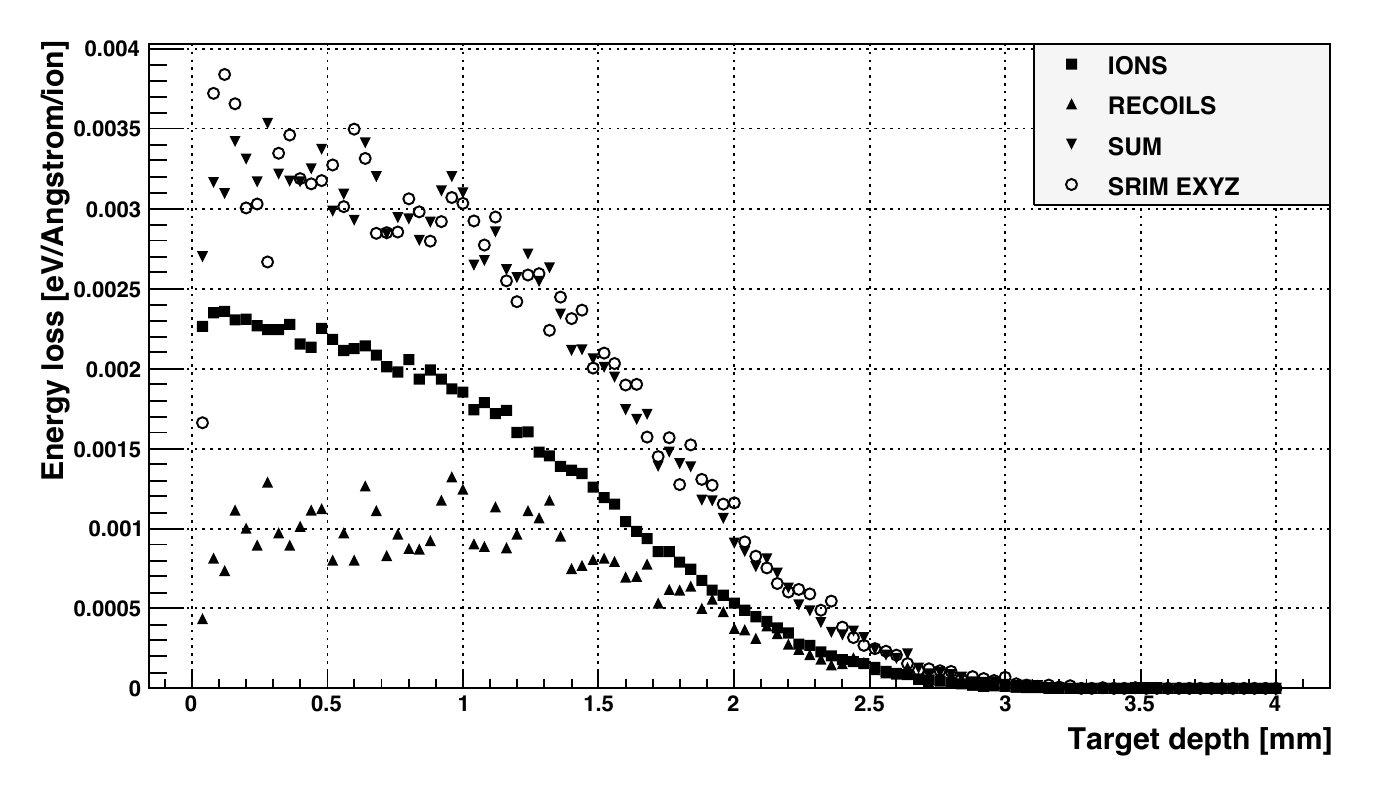}
\caption{ Average ionization energy loss for 100~keV Sulfur ion ($\blacksquare$) and created in the 
cascade secondary Carbon and Sulfur nuclear recoils ($\filledmedtriangleup$) as a function of target 
depth, obtained from TRIM histrogramming. Total energy loss ($\filledmedtriangledown$) is compared 
with scaled results from {\it EXYZ} TRIM output file analysis ($\medcircle$).}
\label{fig8}
\end{center}
\end{figure}

Along with the event-by-event information on ion energy loss, TRIM also 
delivers histograms accumulated on-line and shows distributions of average 
values of several quantities as a function of target depth. With the default 
initial TRIM settings in place the target depth is defined by projection onto 
the axis of the initial direction of ion motion. One of the histograms shows 
distributions of the average ionization energy loss by the primary ion recoils 
and secondary recoils separately as a function of target depth. An example 
of such distributions is shown in Fig. \ref{fig8} for 1000 sulphur ions with energy of 
100~keV. Here one can see the ionization energy loss by the recoil ion, by 
the secondary recoils produced and by their sum. The latter is compared in 
Fig. \ref{fig8} with scaled results derived from analysis of the {\it EXYZ} TRIM output 
file. Reassuringly, both show the same form of ionization energy loss vs. target 
depth, indicating here that more energy is lost at the beginning of the track 
(tail) than at the end (head). 

However, whilst such averaged SRIM plots are applicable for use in, for 
instance, ion damage calculations, caution is needed in interpreting or using 
them directly for studying head-tail discrimination in a dark matter detector. 
This is because the distribution is an average provided as a function of target 
depth regardless of the ion range that, due to straggling, varies from ion to 
ion. For the experiments of interest here, recoils are detected on an event-by-event 
basis with ionization charge distribution created along the track and 
then projected onto an axis. For this situation, as pursued here, use of the 
{\it EXYZ} track generation output file alone is more appropriate. 

\begin{figure}
\begin{center}
\includegraphics{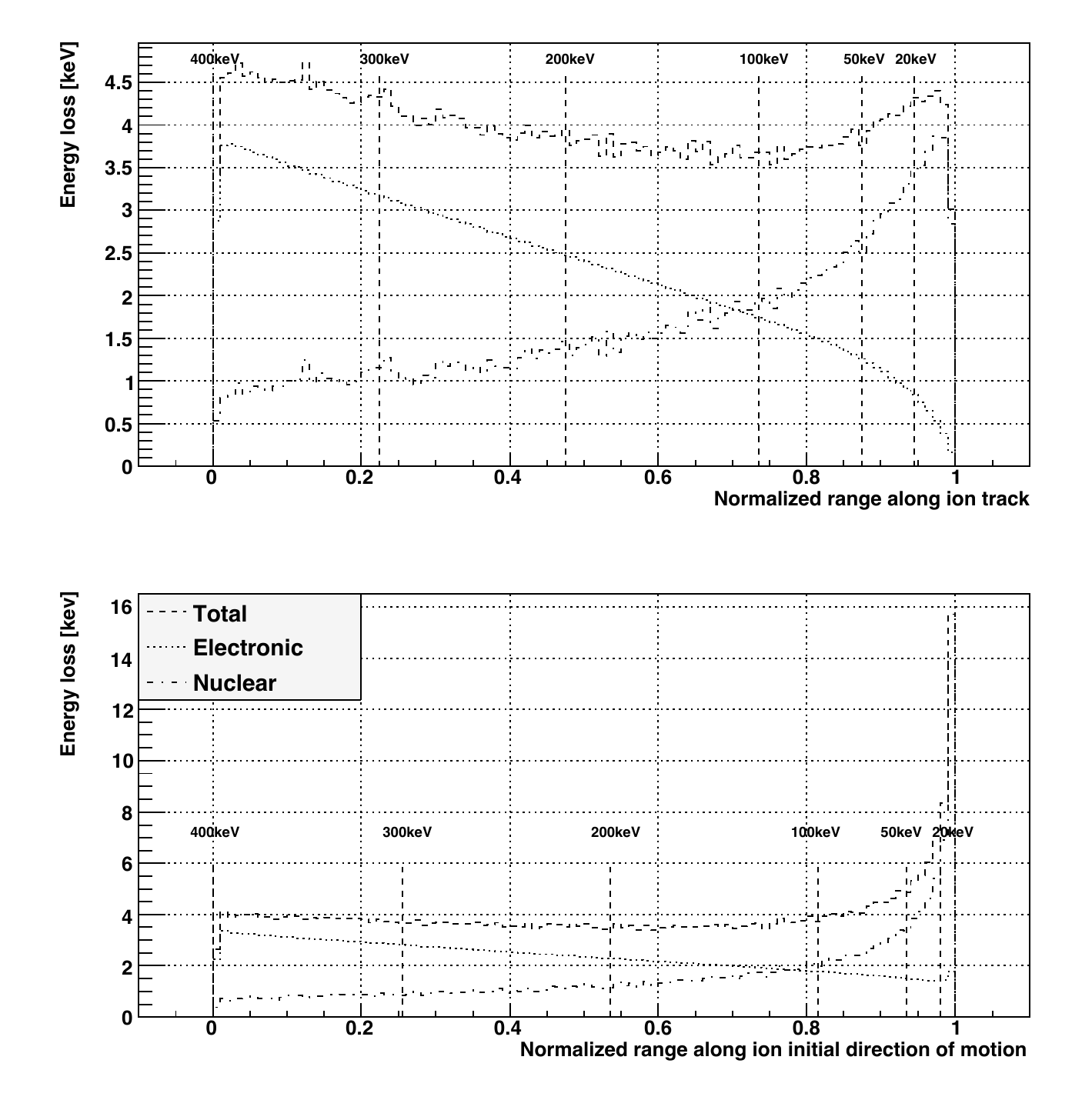}
\caption{Energy loss distribution for 400~keV Sulfur ion along track (top) and along 
direction of  the initial ion motion (bottom). Dashed, dotted and dash-dotted lines 
indicate: total, electronic and nuclear energy loss respectively. Vertical dashed lines 
indicate intermediate ion energy.}
\label{fig9}
\end{center}
\end{figure}

\begin{figure}
\begin{center}
\includegraphics{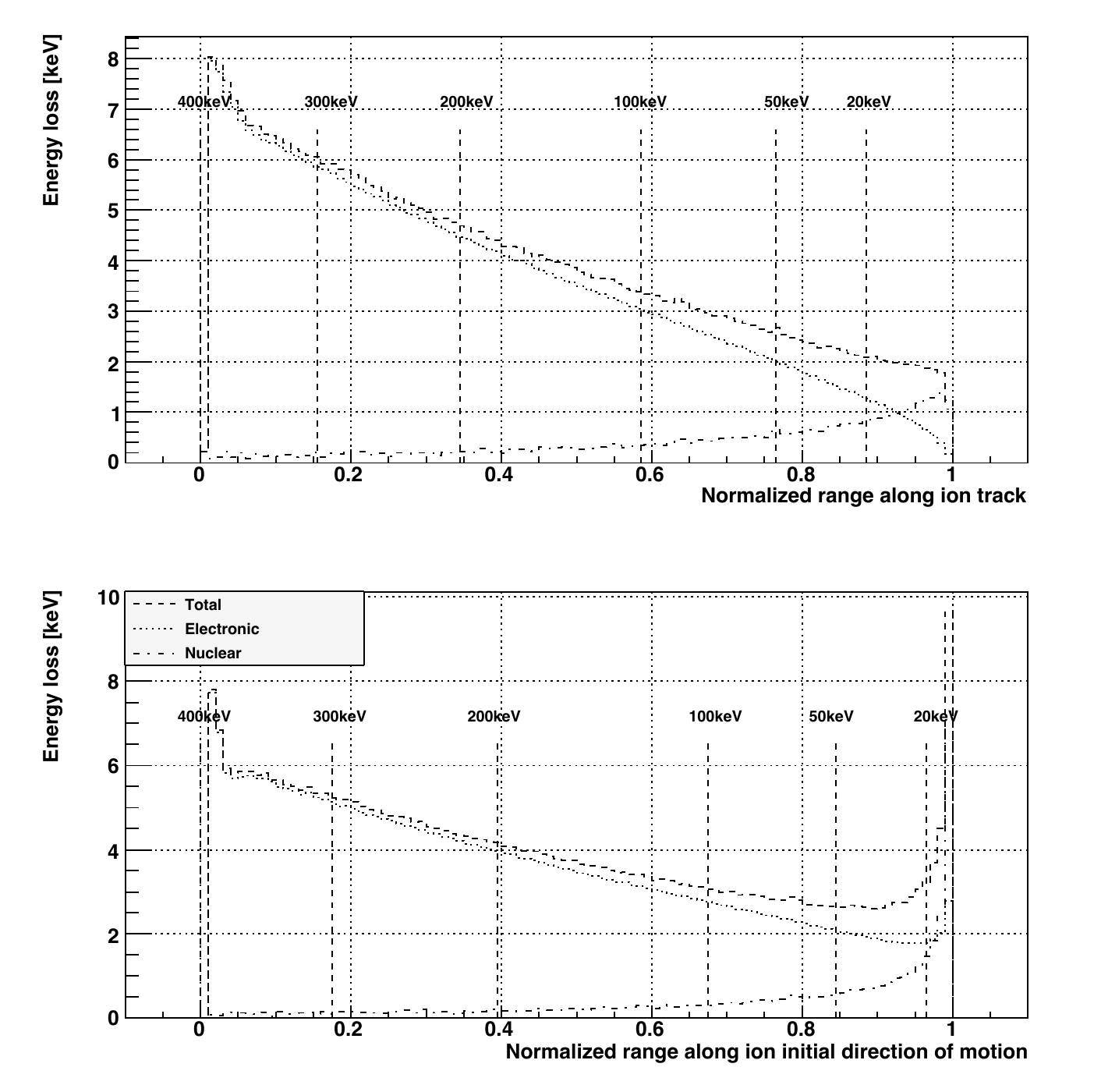}
\caption{As in Fig. \ref{fig9} for Carbon ion.}
\label{fig10}
\end{center}
\end{figure}

\begin{figure}
\begin{center}
\includegraphics{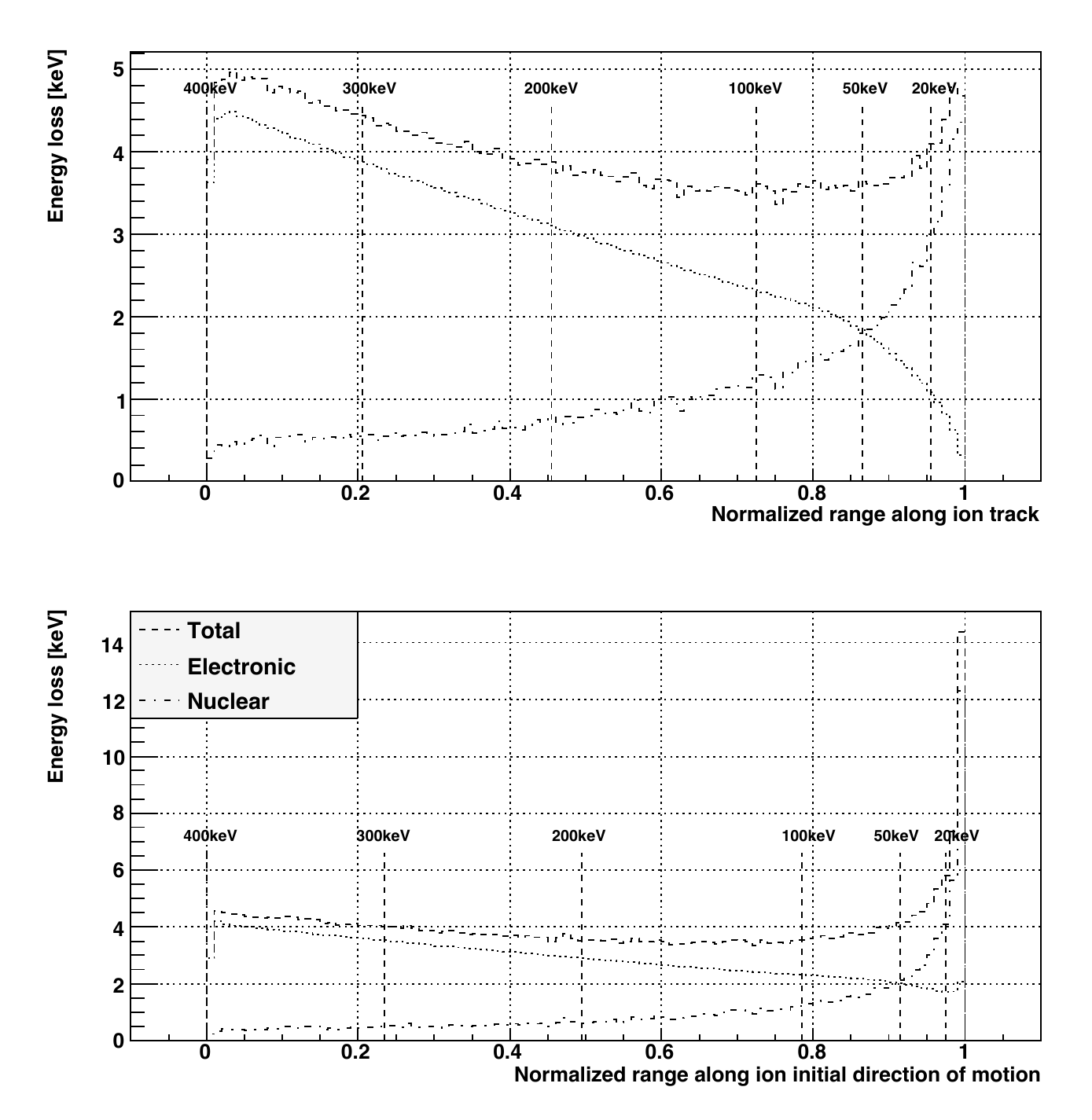}
\caption{As in Fig. \ref{fig9} for Fluorine ion.}
\label{fig11}
\end{center}
\end{figure}

Based on this, Figs \ref{fig9}-\ref{fig11} show the distributions of total, nuclear and 
electronic energy loss for sulphur, carbon and fluorine ions along the track and 
projected onto the initial direction of the ion motion and normalised to its 
projected range R. Calculations were performed for ion energies up to 400~keV. 
As expected from the stopping power shape for both fluorine and carbon ions 
with high energies, where the electronic energy loss dominates, the 
total energy loss distribution has mainly a clear negative slope. This indicates 
greater energy loss at the tail of the track than at the head. However, due to 
ion straggling at the very end of the track, the energy loss there accumulates 
in a small volume that, when projected, creates a clearly visible sharp rise at 
the end of the energy loss distribution. For lower ion energies the slope at the 
beginning of distributions gets smaller and for energies below 200~keV and 
50~keV for fluorine and carbon respectively it becomes flat. Only the sharp end 
remains, indicating now that more energy is lost at the head of the track than 
at the tail. For all studied energies of sulphur ions the energy loss distribution 
remains largely flat but, like the others, ends with a sharp rise. Clearly this 
distribution for sulphur differs from the one given in Fig. \ref{fig8}. It shows an 
opposite effect, because now the influence of straggling is properly included and 
this can dominate over the stopping power characteristic at the head of the 
track. 

Next, converting from energy loss to ionisation (NIPs) charges, the distributions 
of ionization charge were studied for sulphur and carbon ions in CS$_2$. This 
was performed for the geometric scenario when ionization is projected onto the 
axis of the initial ion motion. The charge diffusion in gas was accounted for by 
redistributing the NIPs in space according to the thermal diffusion formula: 
$\sigma=\sqrt{\frac{2k_BT}{e}}\sqrt{\frac{L}{E}}$ where $k_B$ is Boltzmann 
constant, $T$ ambient temperature, $e$ 
electron charge, $E$ drift electric field and $L$ drift distance. After numerical 
simplifications the diffusion formula used was of form: $\sigma[mm] = 0.72\sqrt{\frac{L[m]}{E[keVcm^{-1}]}}$ 
for three different drift lengths $L$: 0.0, 0.3 and 0.5~m and electric field $E$= 0.58~kV/cm as used in the DRIFT IIb detector \cite{ref14}. 

\begin{figure}
\begin{center}
\includegraphics{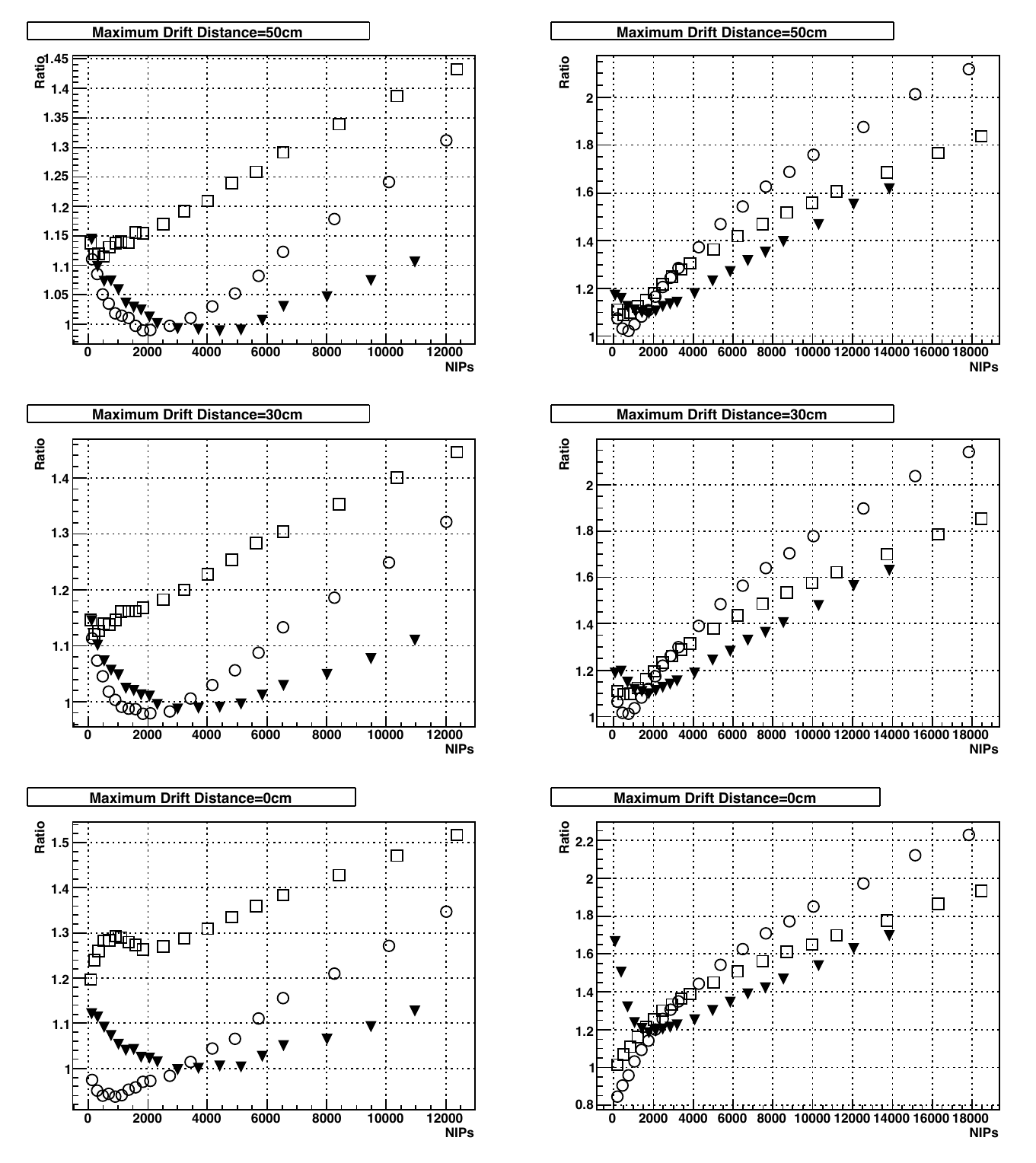}
\caption{Position of the mean of the charge distribution along Sulfur (left column) 
and Carbon (right column) ions tracks in CS$_2$ showing head-tail effect as a function 
of energy and maximum drift distance: 50, 30 and 0~cm, for three different W-values: 
from Hitachi ($\medcircle)$, from Ifft($\filledmedtriangledown$) and electronic ($\medsquare$). 
Points with ratio above 1 indicate 
that more ionization charge is created at the beginning (tail) while those below that 
level, at the end (head) of the track. }
\label{fig12}
\end{center}
\end{figure}

The head-tail effect information was quantified using the same parameter as 
formulated for the DRIFT II data analysis described in \cite{ref7}, on an event-by-event 
basis that is $Ratio=\frac{\int_0^{50\% R}{NIPs}}{\int_{50\% R}^{100\% R}{NIPs}}$
charge integrated along the two halfs
of the projected range R. Fig. \ref{fig12} shows results of the head-tail effect for 
sulphur and carbon tracks, respectively. Results are given as a function of ion 
energy for three different W-value data sets described earlier. To illustrate the 
impact of diffusion for instance as might be expected in an experimental run 
of DRIFT, results are given for tracks drifted for a random distance up to 
maximum of 0, 30 and 50~cm. Points with Ratio values above 1 are for ion 
tracks with greater ÔtailÔ than ÔheadÔ and those below for ions with the reverse. 

Examining these plots it can be seen that again the head-tail effect depends on 
the ion energy as expected. The effect is greatest for carbon at high energies 
where the tail dominates but reduces towards lower energies, the asymmetry 
eventually switching to domination by the head below typically 60~keV. For 
sulphur, the head-tail effect is generally found to be smaller but varies less 
with recoil ion energy and maintains tail domination for almost all energies. 

This is expected since with this geometry the effect of the straggling is 
extracted more accurately. Finally, it can be seen that there is some dependence 
of the head-tail effect on the W-value used. The increased values at low 
energy from the Hitachi et al. prediction result in further enhancement of the 
head-tail asymmetry. 

\section{Conclusions}

In this work preliminary results of the head-tail effect of carbon and sulphur 
ions in low pressure CS$_2$ gas as used in DRIFT detectors have been presented. 
Results were obtained using the SRIM/TRIM package with a purpose-written 
Monte-Carlo and W-values available from both existing and theoretical data. 
The shape of the ionisation charge distribution along an ion track, without 
effect from ion range straggling, is found to be similar to that previously 
presented by Hitachi. It shows that for ion energies down to 10~keV more 
ionization charge is created at the tail than at the head, irrespective of the 
ion type. However, when full account is taken of straggling at the end of 
tracks and diffusion, particularly in relation to the geometry of the readout, 
the situation dramatically changes. For instance, for tracks projected onto the 
direction of the initial ion motion, results indicate the existence of a head-tail
 effect with a greater head for energies below 70 and 350~keV for carbon 
and sulphur respectively. Above these energies the tail starts to dominate. It 
is further found that the head-tail effect depends on the W-value, values for 
which below 20~keV are particularly uncertain. 

In summary, we note two main conclusions from these initial detailed 
simulations. Firstly, that for all the scenarios explored here some form of head-tail 
asymmetry in the low energy ion recoil tracks is expected. Secondly, that the 
nature of this asymmetry, including the sense (whether head or tail is 
dominant), can not be predicted by simple consideration of the average ionisation 
loss simulations, but rather is critically influenced by details of the straggling, 
readout used and W-values at low energy. More experimental effort, underway 
by the DRIFT collaboration and others, is needed to untangle these issues. 

\section{Acknowledgements}

Authors would like to thank prof. A. Hitachi for valuable comments and sharing 
his calculation results of the W-values for Sulfur and Carbon ions in CS$_2$ 
and prof. J.F. Ziegler for his help in SRIM calculations. P. Majewski would 
like to thank support given through University of Sheffeld contract DF7016 
and ILIAS contract RII3-CT-2004-506222.

\clearpage

\begin{table}
\begin{center}
\begin{tabular}{*{3}{|c}|}
\hline
Gas & Measurement [$\mu$m] & SRIM [$\mu$m] \\
\hline
Ar & 79 & 73 \\
\hline
Xe & 44 & 36 \\
\hline
CH$_4$ & 84 & 95 \\ 
\hline
C$_2$H$_4$ & 58 & 61 \\ 
\hline
Air & 83 & 80 \\ 
\hline
N & 80 & 74 \\
\hline 
\end{tabular}
\caption{Ranges of 103 keV 206 Pb recoils in pure and binary gases at STP measured and 
calculated with TRIM. Values from measurement are taken from \cite{ref15}. 1$\sigma$ of the
simulated range distributions is 
10\% of the tabulated mean values for all gases except Xe for which 1$\sigma$ is 15\%. Measurement uncertainty is $\pm$2$\mu$m. }
\label{tab1}
\end{center}
\end{table}

\begin{table}
\begin{center}
\begin{tabular}{*{4}{|c}|}
\hline
Data & a & b & c \\
\hline
S (Ifft) & 32.13 & 289.17 & 0.79 \\ 
\hline
C (Ifft) & 28.78 & 36850.30 & 3.18 \\ 
\hline
S (Hitachi) & -489.4 & 580.46 & 0.02 \\
\hline 
C (Hitachi) & 3.19 & 73.30 & 0.26 \\
\hline
\end{tabular}
\caption{Values of the parameters from the Þt to experimental and theoretical data of the 
W-values shown in Fig. \ref{fig4}}
\label{tab2}
\end{center}
\end{table}

\clearpage

\end{document}